\documentclass[12pt]{article}

\usepackage{scicite}
\usepackage{graphicx}

\usepackage{times}

\renewcommand{\vec}[1]{{\bf #1}}
\newenvironment{sciabstract}{%
\begin{quote} \bf}
{\end{quote}}

\newcounter{lastnote}

\title{Skyrmion Lattice in a Chiral Magnet}

\author
{S. M\"uhlbauer,$^{1,2}$ B. Binz,$^{3}$ F. Jonietz,$^{1}$ C. Pfleiderer,$^{1\ast}$
A. Rosch,$^{3}$\\
A. Neubauer,$^{1}$ R. Georgii,$^{1,2}$ P. B\"oni,$^{1}$\\
\\
\normalsize{$^{1}$Physik Department E21, Technische Universit\"at
M\"unchen, D-85748 Garching, Germany}\\
\normalsize{$^{2}$Forschungsneutronenquelle Heinz Maier-Leibnitz (FRM II)}\\
\normalsize{Technische Universit\"at M\"unchen, D-85748 Garching, Germany}\\
\normalsize{$^{3}$ITP, University of Cologne, Z\"ulpicher Str. 77, D-50937 Cologne, Germany}\\
\\
\normalsize{$^\ast$To whom correspondence should be addressed; E-mail:  christian.pfleiderer@frm2.tum.de.} }

\date{\today}

\begin{document}

\baselineskip24pt

\maketitle

\begin{sciabstract}
Skyrmions represent topologically stable field configurations with particle-like properties. We used neutron scattering to observe the spontaneous formation of a two-dimensional lattice of skyrmion lines, a type of magnetic vortices, in the chiral itinerant-electron magnet MnSi. The skyrmion lattice stabilizes at the border between paramagnetism and long-range helimagnetic order perpendicular to a small applied magnetic field regardless of the direction of the magnetic field relative to the atomic lattice. Our study experimentally establishes magnetic materials lacking inversion symmetry as an arena for new forms of crystalline order composed of topologically stable spin states.
\end{sciabstract}

\newpage
The formation of most crystals is related to three main aspects. First is the interplay of local repulsions and long range attraction of atoms leading to an instability of the liquid with correlations that are completely isotropic and without preferred direction. Second, three particle collisions lower the energy and  dominate the formation of the crystal out of the isotropic density fluctuations. Third, atoms are quantized with an integer number of them in a unit cell.

For the spin structures in magnetic materials it was believed that some or all of the three above mentioned mechanisms could not take place. Consider for instance the formation of a magnetically ordered state out of a paramagnetic metal. The underlying atomic lattice or the Fermi surface strongly breaks the rotational and translational invariance. Furthermore, for magnetic fluctuations in the paramagnet, three particle collisions are forbidden by time reversal symmetry. Lastly, the question arises, as to which quantized entities may play the role of the atoms in a solid. Possible candidates are topologically stable objects such as skyrmions, hedgehogs or merons, which have received great theoretical interest ranging from magnetic monopoles in particle physics \cite{thof74,poly74} to the emergence of gauge theories in condensed matter \cite{levi04}. In analogy to the Abrikosov vortex lattice in superconductors, these objects may be considered as the building blocks of novel forms of electronic order akin to crystal structures. However, the experimental evidence for these building blocks is scarce.

We report the observation of the formation of a magnetic structure with hexagonal symmetry perpendicular to a small applied magnetic field in the cubic B20 compound MnSi. We show that this structure can be described approximately by a simple superposition of three helical states in the presences of a uniform field, where the superimposed state with the lowest energy can approximately be viewed as a lattice of antiskyrmion-lines, that is, magnetic vortices for which the magnetization in the center is antiparallel to the applied field. All three mechanisms mentioned above play an important role in explaining the magnetic structure. The lack of space inversion symmetry in the atomic crystal of MnSi results in weak spin-orbit coupling that generates slow rotations of all magnetic structures such that they decouple from the underlying atomic lattice very efficiently.  Further, in the presence of an external magnetic field, which breaks time-reversal symmetry, three-particle interactions of the magnetic excitations do occur. Lastly, for the magnetic state that emerges under these conditions skyrmion lines, that is, certain topologically protected knots in the magnetic structure, take over the role of the atoms in usual crystals.

At ambient pressure and zero applied magnetic field MnSi develops helical magnetic order below a critical temperature, $T_c=29.5\,{\rm K}$, that is the result of three hierarchical energy scales. The strongest scale is ferromagnetic exchange favoring a uniform spin polarization (spin alignment). The lack of inversion symmetry of the cubic B20 crystal structure results in chiral spin-orbit interactions, which may be described by the rotationally invariant Dzyaloshinsky Moriya (DM) interaction.  The ferromagnetic exchange together with the chiral spin-orbit coupling lead to a rotation of the spins with a periodicity $\lambda_h\approx190\,{\rm \AA}$ that is large compared with the lattice constant, $a \approx 4.56$\,\AA. This large separation of length scales implies an efficient decoupling of the magnetic and atomic structures. Therefore, the alignment of the helical spin spiral along the cubic space diagonal $\langle111\rangle$ is weak and is only fourth power in the small spin-orbit coupling. These crystalline field interactions, which break the rotational symmetry, are by far the weakest scale in the system.

\begin{figure}
\begin{center}
\includegraphics[width=.55\textwidth,clip=]{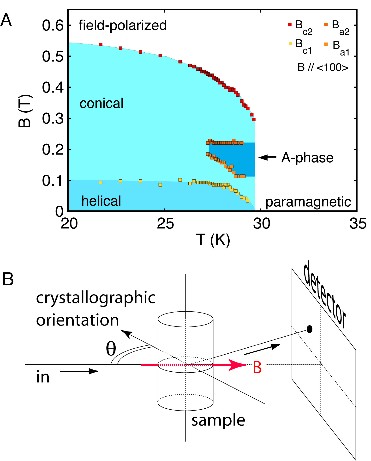}
\end{center}
\caption{(A) Magnetic phase diagram of MnSi. For $B=0$, helimagnetic order develops below $T_c=29.5\,{\rm K}$. Under magnetic field, the helical order unpins and aligns along the field above $B_{c1}$; above $B_{c2}$, the helical modulation collapses. In the conical phase, the helix is aligned parallel to the magnetic field. The transition fields shown here have been inferred from the AC susceptibility, where the DC and AC field were parallel $\langle100\rangle$ \cite{thes97}. (B) Neutron scattering set-up used in our study; the applied magnetic field $\vec{B}$ was parallel to the incident neutron beam.}
\label{figure1}
\end{figure}

Our study was partly inspired by recent work on the pressure dependence of the properties of MnSi \cite{pfle01,pfle04,pfle07}. As shown in Fig.\,\ref{figure1}A well below $T_c$ an applied magnetic field, $\vec{B}$, unpins the helical order and aligns its wave vector $\vec{Q}$ parallel to the field, $\vec{Q}\parallel \vec{B}$, for a field exceeding $B_{c1}\approx0.1\,{\rm T}$ \cite{ishi84,lebe93,grig06b}. This state is referred to as the conical phase. For a magnetic field exceeding $B_{c2}\approx0.6\,{\rm T}$, the effects of the DM interaction are suppressed, giving way to a spin-aligned (ferromagnetic) state. For temperatures just below $T_c$ an additional phase, referred to as the A phase, is stabilized in a finite field interval for $\vec{B}\parallel \langle100\rangle$ (Fig.\,\ref{figure1})\cite{ishi84,lebe93}. It had been believed that the A phase was explained by a single-$\vec{Q}$ helix, where $\vec{Q}$ is perpendicular to the applied field \cite{lebe93}. The specific heat exhibits a tiny peak at the border of the A phase, wheras the AC susceptibility discontinuously assumes a lower value \cite{thes97}.  When taken together with the discontinuous change of the scattering intensity seen in neutron scattering, this suggests that the A phase represents a distinct magnetic phase, where the transition from the conical order to the A phase is discontinuous (first order).

Because $\vec{Q}$ tends to align parallel to an applied magnetic field, neutron scattering as a function of $\vec{B}$ has been reported for set-ups in which the magnetic field was perpendicular to the incident neutron beam \cite{SOM}. In contrast, we chose the incident neutron beam to be parallel to the applied magnetic field (Fig.\,\ref{figure1}B). Two samples were studied. Sample 1 refers to a disk of 19\,mm diameter, $d=3\,{\rm mm}$ thick, where the vector normal to the disc was slightly misaligned by $11^{\circ}$ with respect to a $\langle110\rangle$ axis. Sample 2 is a small parallelepiped, with dimensions $1.5\,{\rm mm}$ by $1.5\,{\rm mm}$ by $14\,{\rm mm}$, where a $\langle110\rangle$ axis corresponded to the long axis. To search for higher order peaks and double scattering we increased the neutron flux in our measurements of sample, accepting a larger beam divergence. In contrast, the beam divergence was reduced for sample 2 to improve the resolution in rocking scans. All data at finite magnetic field were measured after zero-field cooling to the desired temperature, followed by a field ramp to the desired field value. However, the results for the A phase were identical when recorded after field-cooling. For further details of the experimental set-up we refer to \cite{SOM}.

\begin{figure}
\begin{center}
\includegraphics[width=.6\textwidth,clip=]{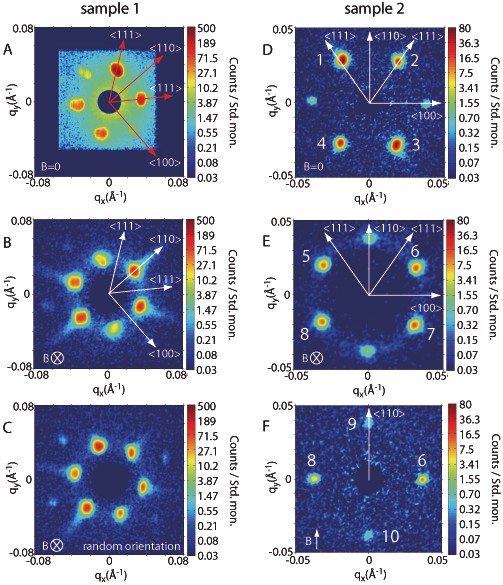}
\end{center}
\caption{Typical neutron small angle scattering intensities; note that the color scale is logarithmic to make weak features visible. Data represent the sum over rocking scans with respect to the vertical axis through the sample. (A) to (C) show data for sample 1 and (D) to (F) for sample 2. Backgrounds measured above $T_c$ and $B=0$ were subtracted in all panels except for (A) (light blue square). Spots are labelled for reference; for the intensity of these spots as a function rocking angle see Ref.\,\cite{SOM}.  (A) Helical order in sample 1 in the zero-field-cooled state  at $T=27\,{\rm K}$ and $B=0$; (B) Six-fold intensity pattern in the A phase in sample 1; same orientation as in (A); $T=26.45\,{\rm K}$, $B=0.164\,{\rm T}$ (C) Six-fold intensity pattern in the A phase for random orientation of sample 1 (see text for details); $T=26.77\,{\rm K}$, $B=0.164\,{\rm T}$ (D) Helical order in sample 2 in the zero field cooled state at $T=16\,{\rm K}$ and $B=0$; (E) A phase in sample 2, same orientation as in (D); $T=27.7\,{\rm K}$, $B=0.162\,{\rm T}$ (F) A phase as measured in conventional set-up (compare Fig.\,1A in \cite{SOM}), where data in all other panels were measured in the configuration shown in Fig.\,\ref{figure1}B); $T=27.7\,{\rm K}$, $B=0.190\,{\rm T}$. A small residual intensity due to the conical phase is observed (spots 9 and 10), whereas spots 6 and 8 correspond to those in (E).}
\label{figure2}
\end{figure}

Fig.\,\ref{figure2} shows typical data recorded, where the spot sizes represent the resolution limit. All data shown represent sums over rocking scans of typically $\pm8^{\circ}$ \cite{SOM}. Fig.\,\ref{figure2}A to C shows data for sample 1, whereas Fig.\,\ref{figure2}D to F shows data for sample 2. Fig.\,\ref{figure2}A shows the scattering intensity of sample 1 in a zero-field-cooled state at a temperature of 27\,K for a $\langle110\rangle$ scattering plane, and the $\langle110\rangle$ axis is indicated in the figure. The pattern is consistent with previously published data and helical magnetic order along $\langle111\rangle$. Fig.\,\ref{figure2}B shows the intensity pattern of sample 1 in the A phase. Six spots emerge on a regular hexagon.

We tested the variation of intensity pattern on the orientation of the field relative to the crystal axes in both samples. The field was always parallel to the incident beam, whereas the sample was rotated for a large number of different orientations. Typical data are shown in Fig.\,\ref{figure2}C for sample 1, where the sample was rotated with respect to the vertical axis into a random position. For both samples and all crystal orientations the scattering pattern always exhibited the six-fold symmetry. In case the scattering plane contained a $\langle110\rangle$ direction, two of the peaks of the six-fold pattern coincided with this direction. As for Fig.\,\ref{figure2}C the scattering plane did not contain a $\langle110\rangle$ direction.  For sample 2, the intensities along the vertical direction, which coincided with the $\langle110\rangle$ direction, were systematically weaker. This may be explained by the demagnetizing fields caused by the large aspect ratio, which implies that part of the scattering intensity was not captured in the rocking scans (see also \cite{SOM}). The main result of our study is that for all orientations of the magnetic field with respect to the atomic lattice six Bragg reflections are observed on a regular hexagon that is strictly perpendicular to the magnetic field.

We performed rocking scans to test whether the A phase has long-range order. Typical data are presented in \cite{SOM}. In the helical state the half-width of the rocking scans corresponded to a magnetic mosaicity $\eta_{m}\approx3.5^{\circ}$ consistent with previous work and long range order \cite{lebe95,pfle07b}. Remarkably, in the A phase the half-width of the rocking scans corresponded to a reduced magnetic mosaicity $\eta_{m}\approx1.75^{\circ}$, implying an even longer correlation length of at least $\xi\approx5500\,{\rm \AA}$, when allowing for demagnetizing fields \cite{SOM}.

To test for consistency with previous work, we have also measured the emergence of the A phase as a function of temperature for magnetic field perpendicular to the neutron beam, where the vertical axis was the same $\langle110\rangle$ axis as before, and the low-symmetry horizontal axis containing spots 6 and 8 in Fig.\,\ref{figure2}F was perpendicular to the magnetic field and incident neutron beam. Data were recorded after: (i) zero-field-cooling the sample to a temperature well below $T_c$, (ii) increasing the magnetic field to 0.19\,T and (iii) measuring the neutron scattering pattern for selected increasing temperatures (Fig.\,\ref{figure2}F shows data for $T=27.7\,{\rm K}$). Well below $T_c$ we first observe the two spots parallel to the field direction labelled 9 and 10 characteristic of the conical state. When entering the A phase the intensity of the spots of the conical phase becomes very weak, but does not vanish, whereas strong scattering intensity appears in the perpendicular direction (spots 6 and 8). This is consistent with previous work and may signal a phase coexistence as expected of a weak first-order transition with possible extra effects of the demagnetizing fields added.

The key results of our neutron scattering data may be summarized as follows: (i) the helical wave-vector aligns perpendicular to the applied magnetic field, (ii) the fundamental symmetry of the intensity pattern is sixfold suggesting a multi-$\vec{Q}$ structure, and (iii) the A phase stabilizes in a magnetic field strength of order $B_{c2}/2$. Moreover, the pattern aligns very weakly with respect to the $\langle110\rangle$ orientation. We can readily account for these features in the framework of standard Landau-Ginzburg theory by taking thermal fluctuations into account on top of a mean-field approximation. Near $T_c$ the Ginzburg-Landau energy functional can be written as \cite{naka80,bak80}
\begin{equation}
F[\vec M]=\int\!d^3r\,\left( r_0\vec M^2+J(\nabla\vec M)^2+2D\,\vec
M\cdot(\nabla\times\vec M)+U\vec M^4 -\vec B\cdot \vec M\right),
 \label{GL}
\end{equation}
The first and second terms represent the usual quadratic contribution with the conventional gradient term, the third term the Dzyaloshinsky-Moriya interaction and the last term the coupling to an external magnetic field $\vec B$. The quartic term accounts in lowest order for the effects of mode-mode interactions and stabilizes the magnetic order. We neglect higher order spin-orbit coupling terms describing anisotropy effects \cite{naka80,bak80}. The free energy is given by $\exp(- G)=\int\!{\cal D}\vec M\,\exp(-F[\vec M])$ (throughout the paper, we use a dimensionless free energy). Within mean field approximation $G(\vec B)$ is roughly equal to $\mbox{min} F[\vec M]$, and one minimizes $F$ with  respect to the spin structure ${\vec M(\vec r)}$.

To explain the A phase, we evoke strong analogies with the crystal formation of ordinary solids out of the liquid state. The latter is in most cases driven by the cubic interactions of density waves\cite{chai95}, which in momentum space can be written as
$$\sum_{\vec{q}_1, \vec{q}_2, \vec{q}_3} \rho_{\vec{q}_1}
\rho_{\vec{q}_2} \rho_{\vec{q}_3}
\delta(\vec{q}_1+\vec{q}_2+\vec{q}_3).$$
The ordered state can gain energy from this term only when three ordering vectors of the crystal structure add up to zero. Accordingly, in many cases (exceptions can arise only for strong first order transitions) the ordered phase which forms first out of a liquid state is of body-centered cubic symmetry \cite{chai95}, which is the crystal structure with the largest number of such triples of reciprocal lattice vectors.

\begin{figure}
\begin{center}
\includegraphics[width=.9\textwidth,clip=]{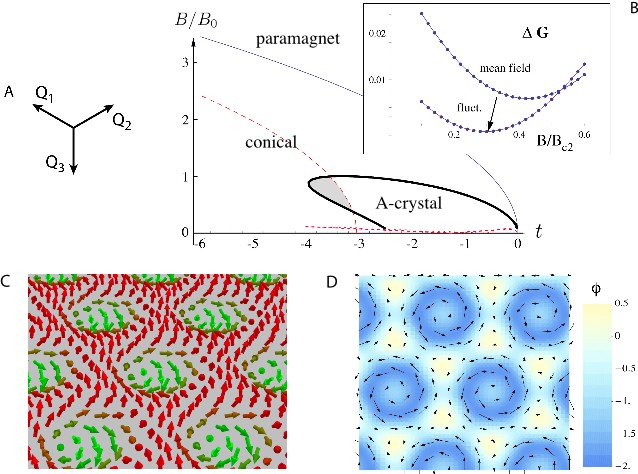}
\end{center}
\caption{ (A) Depiction of the hexagonal basis vectors of the crystalline spin order in the A phase. (B) Theoretical phase diagram as a function of magnetic field $B/B_0$ with $B_0=\sqrt{(J Q^2)^3/U}$ and the parameter $t=r_0J/D^2-1$, which is roughly proportional to $T-T_c$.  We use the model parameter $\gamma=JD/U=5$ and a momentum-space cutoff $k<40 D/J$. Smaller values of $\gamma$ increase the A phase regime. For most values of field, the A phase is either metastable or stable, but at low fields below the dotted line, it becomes unstable. Above and to the right of the red dashed line, fluctuation correction to the size of the order parameter becomes larger than 20\% and our theoretical analysis becomes uncontrolled. Therefore in the shaded gray region, we have  therefore reliably established stability of the A phase within our model (see \cite{SOM} for details). (Inset) Energy difference between A phase and conical phase as a function of field for the same parameters and $t=-3.5$, both in the mean-field approximation and with fluctuation corrections. Fluctuations stabilize the A phase at intermediate fields. (C) Real space depiction of the spin arrangement in the A phase  in the $x$-$y$ plane. Note that this spin arrangement is translation-invariant along the z-axis, which is parallel to the magnetic field. (D) Skyrmion density per unit cell area as calculated for the A phase as shown in panel (C). The integrated skyrmion density per unit cell is finite, $\Phi=-1$. The arrows represent the magnetization component perpendicular to the line of sight.} 
\label{figure4}
\end{figure}

In the presence of a finite uniform component of the magnetization, $\vec{M}_f$, a similar mechanism can also occur in MnSi.  From the quartic term in Eq.\,\ref{GL} we obtain terms which are effectively cubic in the modulated moment amplitudes
\begin{equation} \sum_{\vec{q}_1,\vec{q}_2,\vec{q}_3} (\vec{M}_f \cdot \vec{m}_{\vec{q}_1})
(\vec{m}_{\vec{q}_2}\cdot \vec{m}_{\vec{q}_3}) \delta(\vec{q}_1+\vec{q}_2+\vec{q}_3),\label{cubic}
\end{equation}
where $\vec{m}_{\vec{q}}$ is the Fourier transform of $\vec{M}(\vec r)$. As in the case of an ordinary crystal, one can gain energy from this term for a structure with three $\vec{Q}$-vectors adding up to zero. These vectors have a fixed modulus -- determined by the interplay  of the two gradient terms in Eq.~\ref{GL}. Therefore these three vectors have relative angles of $120^\circ$ (Fig.\,\ref{figure4}A) and define a plane characterized by a normal vector $\hat{n}$. By symmetry, the energy change is proportional to $\vec{M}_f \cdot \hat{n}$ and therefore the three $\vec{Q}$ vectors must be perpendicular with respect to the external magnetic field. Our qualitative arguments already explain the two main experimental observations in the A phase: The Bragg spots are located in the plane perpendicular to $\vec{B}$ and display a sixfold symmetry (because both $\vec{Q}$ and $-\vec{Q}$ give a Bragg reflection) independent of the orientation of the underlying lattice. We therefore suggest that the A phase is a chiral spin crystal, the A crystal, approximately characterized by the magnetization
\begin{equation}\label{aphase}
\vec{M}(\vec{r})\approx  \vec{M}_f+ \sum_{i=1}^3
\vec{M}^h_{\vec{Q}_i}(\vec{r}+\Delta \vec{r}_i)
\end{equation}
where
$\vec{M}^h_{\vec{Q}_i}(\vec{r})=A (\vec{n}_{i1} \cos(\vec{Q}_i
\vec{r})+\vec{n}_{i2} \sin(\vec{Q}_i)
\vec{r}) $ is the magnetization of
a single chiral helix with amplitude $A$, wave vector $\vec{Q}_i$ and
two unit vectors, $\vec{n}_{i1}$ and $\vec{n}_{i2}$, orthogonal to each other and to $\vec{Q}_i$.
All three helices have the same chirality, i.e., all
$\vec{Q}_i\cdot (\vec{n}_{i1}\times \vec{n}_{i2})$ have the same sign.
More precisely, one has to add further higher-order Fourier
components to Eq.~\ref{aphase} when minimizing $F[M]$. However, these terms
remain small close to $T_c$.
The relative shifts $\Delta \vec{r}_i$ of the helices, which
we calculate theoretically, determine whether the A crystal can be
described as a lattice of skyrmions (see below).

One also has to take into account, that an external magnetic field
favors helices with $\vec{Q}$ vectors parallel to $\vec{B}$, because
this way the spins may easily tilt parallel to the field to form a
conical structure. Within mean-field theory, this conical phase always has
the lowest energy \cite{SOM}. However, in the parameter range,
where the A phase occurs experimentally, i.e., close to $T_c$ at
intermediate magnetic fields (Fig.~\ref{figure1}A), the energy
difference between the two phases becomes very small as shown in 
Fig.\,\ref{figure4}B inset.  The origin of the energy minimum of
the A crystal for moderate magnetic field can be traced back to the
size of the modulations of the magnetization amplitude,
$|\vec{M}(\vec{r})|$ which is minimal close to $B\approx 0.4 B_{c2}$
\cite{SOM}). In our mean-field Landau-Ginzburg theory, the A crystal
thus appears as a meta-stable phase, which becomes extremely close in
energy to the conical phase for intermediate fields $B\approx 0.4
B_{c2}$.

It turns out that when we consider thermal fluctuations around the mean field solution these stabilize the A crystal. To show this, we consider the  leading correction to mean field theory arising from Gaussian fluctuations
\begin{equation}
G\approx F[\vec
M_0]+\frac12\log\det\left.\left(\frac{\delta^2F}{\delta\vec M\delta
      \vec M}\right)\right|_{\vec M_0},\label{sp}
\end{equation}
where $\vec M_0$ is the mean field spin configuration for either the
A phase or the conical phase. To make Eq.\,\ref{sp} well defined,
one has to specify a cutoff scheme for short length scales. We use a
cutoff in momentum space, $k<2 \pi /a$, where $a$ is the lattice
spacing of the MnSi crystal. Because the long pitch of the helix, it
turns out that most contributions arise from fluctuations on short
length scales with the exception of temperatures extremely close to
$T_c$ (see \cite{SOM} for a detailed discussion), 
but both short-range and long-range fluctuations favour
the A crystal for intermediate magnetic fields.  As shown in 
Fig.~\ref{figure4}B inset, the fluctuations indeed stabilize the
A crystal. A typical phase diagram resulting from (\ref{sp}) is shown
in Fig.~\ref{figure4}B. The theoretical phase diagram catches the main
characteristics of the experimental phase diagram. The A crystal is
stable at intermediate fields not too far from $T_c$. When
interpreting the theoretical result one has to take into account that
Eq.\,\ref{sp} is only valid for small fluctuations and therefore can
not be applied too close to $T_c$. Indeed it is expected
\cite{braz75,schm04} that fluctuations ultimately drive the transition
first order and that such strong fluctuations will substantially shift the
transition line to the paramagnet. We estimate the strength of
fluctuations by calculating the leading correction to the order
parameter for both the conical phase and the A crystal \cite{SOM}. In
the shaded area of Fig.~\ref{figure4}B these corrections are small
(less than 20\%), which justifies the use of Eq.~\ref{sp}.

As can be seen from Fig.~\ref{figure4}C, the magnetic structure of the
A crystal obtained by minimizing $F[\vec{M}]$ is characterized by a pattern of magnetic vortices. To elucidate their nature we compute the skyrmion density given by
\cite{binz08}:
\begin{equation}
\phi=\frac{1}{4 \pi} \vec{n}\cdot \frac{\partial \vec{n}}{\partial x}\times \frac{\partial \vec{n}}{\partial y}
\end{equation}
where $x$ and $y$ are the coordinates perpendicular to $\vec{B}$ and $\vec{n}=\vec{M(\vec{r}})/|\vec{M(\vec{r})}|$ is the orientation of the magnetization. $\phi$ is a measure of the winding of the magnetization profile. If $\phi$ integrates to $1$ or $-1$, a topologically stable knot exists in the magnetization. As shown in Fig.~\ref{figure4}D, the skyrmion density is finite and oscillates between positive and negative as compared with the normal helical or conical phases, where it is zero. Moreover, the skyrmion number $\Phi=\int \phi(\vec{r}) d^2 \vec{r}$ per two-dimensional unit cell is quantized and adds up to $-1$. Taken together this implies that the A crystal can be interpreted as a crystal made out of quantized objects, the skyrmion lines, with a magnetization at their core that is antiparallel to the applied magnetic field and $\vec{M}_0$.

Because of its symmetry and the cubic interactions, the A crystal has to be separated by two first order phase transitions both from the conical and the paramagnetic phases. This is consistent with the experimental observations. Two additional features in the scattering patterns that account for less than 1\%  of the total integrated scattering intensity are, first, weak continuous streaks of intensity emerging radially outward from the six main spots, and second, the coexistence of conical order and the spin crystal. Both may be the result of weak heterogeneities resulting from these generic first-order boundaries of the A crystal; possibly in combination with demagnetizing fields.

Many years ago, Bogdanov and collaborators used a mean-field model to establish the existence of skyrmion lattices for anisotropic non-centrosymmetric magnetic materials under the application of a magnetic field \cite{bogd94,bogd89}. The authors also pointed out that within their mean-field analysis for cubic materials such as MnSi the skyrmion lattice is always metastable.  Moreover, in the absence of a magnetic field, it has been shown that certain crystalline spin structures can be stabilized by long-range interactions or an additional phenomenological parameter \cite{binz06,roes06,fisc08}. In contrast to this work, we have reported here that it is sufficient to include the effects of Gaussian thermal fluctuations to stabilize skyrmion lattices in a magnetic field in cubic materials.

It is instructive to search for analogies of the A crystal in other condensed matter systems. Because it is a multi-$Q$ structure we note that previously known multi-$Q$ structures, for example, in the rare earths involve large values of $Q$ and exhibit very strong pinning to the atomic lattice \cite{forg89,gign91}, whereas for MnSi we observe that the six-fold pattern of the A crystal exists independently of the underlying lattice and $Q$ is quite tiny. Although flux lines in superconductors and the magnetic skyrmion lines observed are topologically completely different objects, there is nevertheless an intimate similarity of the Abrikosov lattice of superconducting flux lines and the hexagonal symmetry of the A crystal \cite{bogd94,bogd89}. Moreover, the A crystal is characterized by broken translation symmetry in the plane perpendicular to $\vec{B}$ only. Therefore the A phase is similar to the chiral columnar phase of liquid crystals \cite{chai95,grel08}. Further, the spin structure of the A crystal is topologically equivalent to theoretical predictions of the spin structure of the ferromagnetic quantum Hall state near 1/2-filling \cite{brey95}, where, however, the underlying energetics is completely different. Lastly, individual magnetic vortices attract also great interest as a micromagnetic phenomenon, which arises when conventional domain walls in soft ferromagnets are made to meet \cite{bade06}.

The skyrmion lattice in the chiral magnet MnSi reported here represents an example where an electronic liquid forms a spin crystal made from topologically  non-trivial entities. This provides a glimpse of the large variety of magnetic states that may be expected from the particle-like magnetic objects currently discussed in the literature \cite{acknow}.

\newpage


\newpage

\newpage

\newpage

\end{document}




\baselineskip24pt


\maketitle






\newpage
\section{Experimental details}

Our neutron scattering measurements were performed at the diffractometer MIRA at FRM II at the Technische
Universit\"at M\"unchen \cite{geor07}. Data were recorded for an incident neutron wavelength $\lambda=9.6\,{\rm
\AA}$ with a 5\% FWHM wavelength spread. A delayline $^3$He area detector of $200\times200\,{\rm mm^2}$ was used
with a position resolution of order $2\times2\,{\rm mm^2}$. The neutron beam was collimated over a distance of 1.5\,m. For sample 1 the source aperture and the aperture at the sample were both $8\times8\,{\rm mm^2}$. For sample 2 the source aperture and the aperture at the sample were both $4\times4\,{\rm mm^2}$. However, the size of sample 2 of $1.5\times4\,{\rm mm^2}$ acted as an effective smaller aperture at the sample reducing the beam divergence even further. The distance between the sample and the detector was between 0.8 and $1.3\,{\rm m}$.

Samples were cooled with a cryogen free pulse tube cooler. The magnetic field was generated with bespoke water-cooled Cu solenoids in a Helmholtz configuration \cite{mueh05}. The magnetic field profile was carefully characterized with a Hall probe and found to be uniform much better than 1\% over the sample volume. All data at finite magnetic field were measured after zero-field cooling to the desired temperature, followed by a field ramp to the desired field value. However, in the A-phase data were identical also after field-cooling. 

\begin{figure*}
\begin{center}
\includegraphics[width=.65\textwidth,clip=]{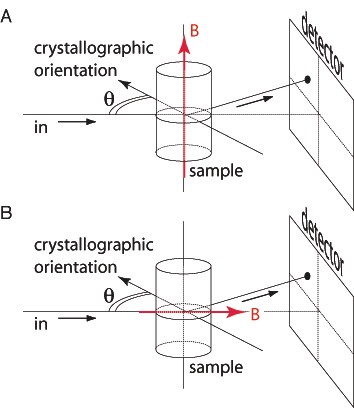}
\end{center}
\caption{(A) Neutron scattering set-up used in all previous studies to explore the precise magnetic structure as function of magnetic field. Note that $\vec{B}$ is perpendicular to the incident neutron beam. (B) Neutron scattering set-up used in our study; the applied magnetic field $\vec{B}$ was parallel to the incident neutron beam.} \label{figure-S0}
\end{figure*}

Because the ordering wave-vector $\vec{Q}$ in MnSi over large portions of the magnetic phase diagram tends to align parallel to an applied magnetic field, neutron scattering as a function of $\vec{B}$ has been reported for set-ups where the magnetic field was perpendicular to the incident neutron beam as shown in Fig.~\ref{figure-S0}A. In contrast, we chose the incident neutron beam to be parallel to the applied magnetic field as shown in Fig.~\ref{figure-S0}B.

Two samples were studied. Sample 1 refers to a disk of 19\,mm diameter, $d=3\,{\rm mm}$ thick, where the vector normal to the disc was slightly misaligned by $11^{\circ}$ with respect to a $\langle110\rangle$ axis. Sample 2 refers to a small parallepiped, with dimensions $1.5\,{\rm mm} \times 1.5\,{\rm mm} \times 14\,{\rm mm}$, where a $\langle110\rangle$ axis corresponded to the long axis.

\section{Demagnetizing fields}

In our study we find that the hexagonal magnetic scattering intensity in the A-phase aligns strictly perpendicular to the applied magnetic field. This implies that the magnetic scattering intensity depends sensitively on the distribution of magnetic field directions across the sample volume, notably the effects of demagnetizing fields.

\begin{figure}
\begin{center}
\includegraphics[width=.65\textwidth,clip=]{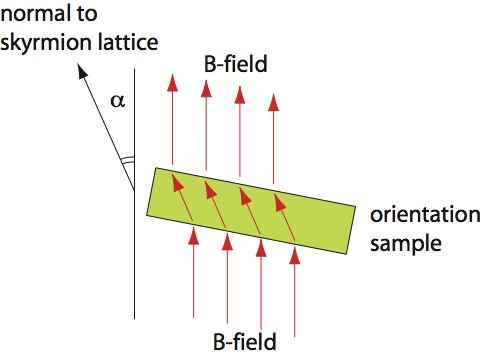}
\end{center}
\caption{Sketch of the deflection $\alpha$ of the orientation of the skyrmion lattice with respect to the sample orientation and magnetic field as observed in sample 1. This deviation is in perfect agreement with the effects of demagnetizing fields.} \label{figure-S1}
\end{figure}

The perhaps most pronounced consequence of the demagnetizing fields is seen in the orientation of the hexagonal scattering intensity with respect to the orientation of sample 1. As stated above this sample is a circular disc of 19\,mm diameter, $d=3\,{\rm mm}$ thick, where the vector normal to the disc was slightly misaligned by
$11^{\circ}$ with respect to a $\langle110\rangle$ axis. When the applied field was not perfectly perpendicular to the disc, we observed a deflection of the maxima consistent with the demagnetizing fields as shown in Fig.~\ref{figure-S1}.

\section{Rocking scans}

In our study rocking scans could only be carried out for the vertical axis. This is particularly important for sample 2, which was a bar with a cross-section of $1.5\times1.5\,{\rm mm^2}$ and 14\,mm long. The long side of sample 2 was parallel to a $\langle110\rangle$ direction. In the neutron scattering experiment the long side of the sample was vertical and perpendicular to the incident neutron beam and magnetic field (cf set-up in Fig.~\ref{figure-S0}B). Thus in the A-phase two spots of the hexagonal scattering intensity coincided always with the vertical $\langle110\rangle$ axis and thus the axis of the rocking scans (cf Fig.\,2E in the main
text).

We note that for a reasonably well collimated neutron beam with a divergence smaller than the rocking width, the observation of equal intensity of the vertical spots implies, that neither of these vertical spots satisfies the Bragg condition. The intensity corresponds only to the tails of the Bragg spot. Therefore, when summing over
rocking scans with respect to the vertical axis, these vertical spots remain weak.  In other words the scattering intensity in the vertical spots was not fully captured in rocking scans, with possible small effects of demagnetizing fields added.

We have also observed weak intensity at the positions of higher-order peaks, although this may arise from double scattering rather than true higher order reflections. An example of double scattering is the appearance of weak spots along $\langle100\rangle$ in Fig.\,2A of the main text as expected of a multi-domain state of helical order at $B=0$. As shown Fig.\,2B and C of the main text to achieve a higher flux and thus larger intensity in sample 1, which had a larger sample volume, we accepted a larger beam divergence. This revealed additional weak spots for the A-phase. However, because the overall intensity was still very weak, it was not possible to distinguish unambiguously if these spots were due to double scattering alone or whether they represented higher order peaks.

\begin{figure}
\begin{center}
\includegraphics[width=.6\textwidth,clip=]{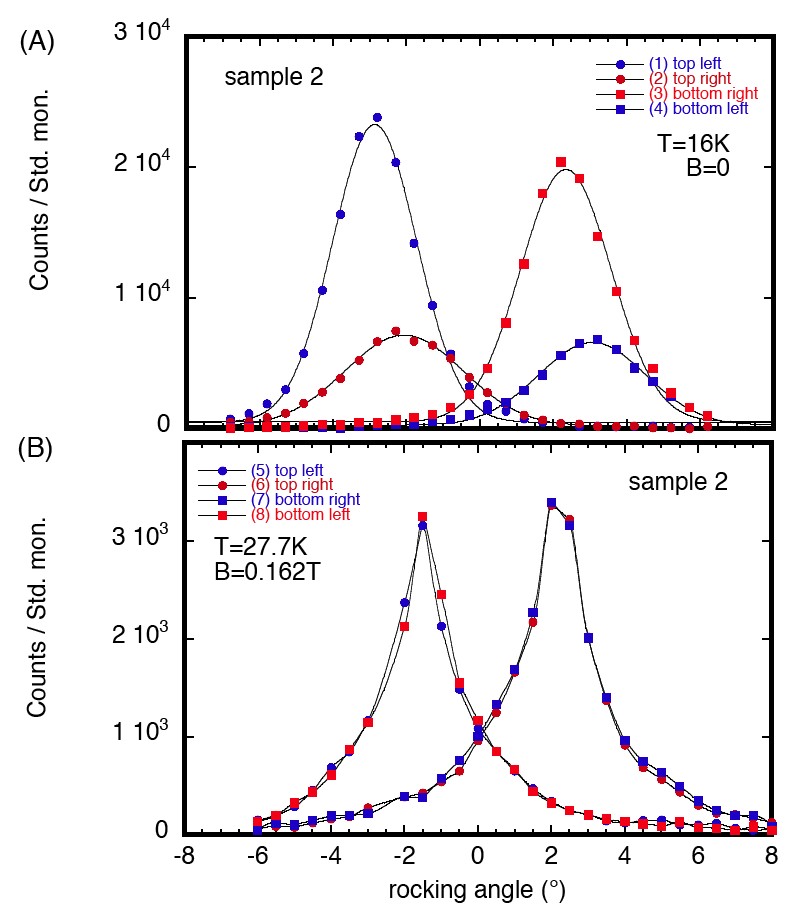}
\end{center}
\caption{Typical variation of the scattering intensity as recorded in rocking scans of sample 2. Intensity represents integrated counts over the individual spots, were labels refer to Fig.\,2 in the main text. Lines serve as to guide the eye. (A) Intensity in the zero-field-cooled state at $T=16\,{\rm K}$ and $B=0$. A spontaneous difference of intensity between the domain populations (1,3) and (2,4) is observed. The half-width
of the rocking angle implies a magnetic mosaicity $\eta_{m}\approx 3.5^{\circ}$ consistent with previous work \cite{harr95,pfle07b}. (B) Intensity in the A-phase on a linear scale for spots (5), (6), (7) and (8). The half-width of the rocking angle implies a magnetic mosaicity $\eta_{m}\approx 1.75^{\circ}$ when taking into account the Lorentz factor of 15\% for the location of the spots, corresponding to a coherence length $\xi\approx5500\,{\rm \AA}\pm10\%$ characteristic of long range order.} \label{figure3}
\end{figure}

While the search for higher order peaks was inconclusive, it was possible to infer evidence of long range order from the rocking scans shown in Fig.~\ref{figure3}. In the helical state the half-width of the rocking scans corresponded to a magnetic mosaicity $\eta_{m}\approx3.5^{\circ}$ consistent with previous work and long range
order \cite{pfle07b}. Remarkably, in the A-phase the half-width of the rocking scans corresponded to a reduced magnetic mosaicity $\eta_{m}\approx1.75^{\circ}$, implying an even longer correlation length of at least $\xi\approx5500\,{\rm \AA}$ when allowing for demagnetizing fields, where the Lorentz factor was taken into
account. A more refined analysis of the peak shape, e.g., considering issues discussed in Ref.\,\cite{lave08},
is beyond the present study.

\newpage

\section{Theoretical model}

In the following we give a detailed description of our theoretical
calculations where we show that thermal fluctuations can stabilize the
skyrmion crystal. We begin with a few remarks of previous work on
skyrmion lattices in Section \ref{remarks}. Due to the long and rich
history of the subject in fields ranging from high-energy physics to
soft matter the discussion will unfortunately remain incomplete. We
next revisit the standard phenomenological Landau-Ginzburg model for
helimagnets such as MnSi in an applied magnetic field. We analyze this
model taking into account Gaussian fluctuations on top of the
mean-field solution. Model and approximation scheme are introduced in
Section \ref{definition}. The well-known mean-field ground state of
this model is the conical phase. In Section \ref{intuition}, we give
simple arguments that for intermediate fields, there is a competing
ground state, which corresponds to a triangular spin crystal of
skyrmions (the A-crystal). In Section \ref{mean-field}, we consider
the mean-field approximation and find that while the conical phase
remains the mean-field ground state, the energy of the A-crystal comes
very close at intermediate fields. We then show in Section
\ref{fluctuations} that the fluctuation corrections to the free energy
lower the energy of the A-crystal compared to the conical state and
make it stable. Finally, in Section \ref{pinning}, we comment on the
experimentally observed very weak pinning of the magnetic A-crystal to
the atomic crystal.

\subsection{General remarks}
\label{remarks}

Continuous fields with topologically non-trivial, stable,
particle-like properties have attracted great interest in a large
number of disciplines for many decades (see also \cite{chai95}). Many
of these studies have been inspired by the Skyrme's seminal work \cite{skyr62}. He
was able to show that in a certain field theory describing
interacting mesons one can find quantized and topologically stable
field configuration. Surprisingly these excitations made of bosonic fields can
be interpreted as baryons, i.e. as fermions.
In the area of nuclear and particle physics, a
very large body of work has been published related to Skyrme's
work. Regarding the formation of skyrmion lattices and related
structures it has, for instance, been suggested that dense nuclear
matter in neutron stars may be described by a three-dimensional
lattice of skyrmions \cite{kleb85}, where typical face-centered cubic
structures give the lowest energy \cite{kugl88,cast89}.

Skyrmion lattices and related structures have also been considered in
many other areas of physics. A long time ago, Bogdanov and coworkers
\cite{bogd89} showed for non-centrosymmetric magnetic materials
that chiral interactions represent an elegant route to skyrmion ground
states \cite{bogd89,bogd95}. In these studies it was found theoretically
that for a class of anisotropic systems skyrmion lattices are the mean-field
ground states for a range of parameters. 
Experimental examples confirming these predictions remain to be
discovered. The authors also pointed out the similarity
of these states to vortex lattices in superconductors. Indeed, it was
emphasized by Bogdanov and Hubert \cite{bogd94}, that a magnetic field can
be instrumental in stabilizing skyrmion lattices. The
authors also pointed out that, within their mean-field analysis 
for cubic materials like MnSi, the energy of a conical state is \textit{always}
lower than the one of  skyrmion lattices as is discussed in
Section~\ref{mean-field} below. One way to stabilize crystalline magnetic
structures, e.g., the skyrmion lattice, for cubic systems on the mean-field level is to consider long-ranged interactions \cite{tewa06,fisc08} or extra
phenomenological parameters \cite{roes06} added to the conventional
Ginzburg-Landau approach. Instead, we will show below, that it is sufficient to
include the effects of Gaussian thermal fluctuations to stabilize
skyrmion lattices in a magnetic field.

A large body of literature on skyrmion lattices is also available in
the area of quantum Hall systems. For instance, the two-dimensional
analog of the Skyrme model can be used to describe excitations in
ferromagnetic quantum Hall systems with small Zeemann splitting where
the spin-density takes over the role of the pion field
\cite{sond93}. In these systems the quantized winding number of the
skyrmion can be identified with the charge of the excitation. Several
theoretical proposals suggested also lattice states of these
two-dimensional skyrmions \cite{brey95,timm98,gree00}. Nuclear
magnetic resonance measurements \cite{desr02,gerv05} and recent
inelastic light scattering experiments \cite{gall08,hen08} gave
results consistent with the existence of such a skyrmion lattice in
GaAs heterostructures with fillings close to $\nu=1$, but there is no
direct microscopic evidence.

The physics of magnets without inversion
symmetry finally shares many similarities with cholesteric liquid crystals. In
these cholesterics, a sequence of several phase transitions from
helical to so-called 'blue phases' has been observed. Blue phases are
characterized by complex patterns of order parameters woven from
topological defects of the underlying chiral helices \cite{merm89}
which appear as colorful objects when typical lattice distances are of
the order of the wave length of visible light. Skyrmion textures have
also been considered in these systems \cite{bogd03}.

\subsection{Definition of the model and saddle point approximation}
\label{definition}

It has long been established \cite{naka80,bak80}, that the helimagnetism in MnSi is well described by the following leading-order Landau-Ginzburg functional, which depends on the continuously varying magnetization $\vec M$: \be F[\vec M]=\int\!d^3r\,\left( r_0\vec M^2+J(\nabla\vec M)^2+2D\,\vec M\cdot(\nabla\times\vec
M)+U\vec M^4 -\vec B\cdot \vec M\right), \label{F} \ee where $\vec B$ is the external magnetic field and $r_0,J,D,U$ are parameters ($U,J>0$). Here, we chose $D>0$, which selects a left-handed spiral with wavevector $Q=|\vec Q|=D/J$. Within the Landau-Ginzburg approach, one linearizes all $T$ dependences around $T_c$. We
therefore keep only a linear $T$-dependence of $r_0$ (see below).

The (dimensionless) free energy $G$ as a function of magnetic field and temperature is obtained as 
\be
e^{-G}=\int\!{\cal D}\vec M\,e^{-F[\vec M]}.\label{G} 
\ee 
To evaluate the functional integral in Eq.~(\ref{G}), we use the saddle point approximation (method of the steepest descent, Laplace's method), which consists of expanding $F$ around its local minima and performing Gaussian integrals over fluctuations around these minima\cite{chai95}. For a given local minimum $\vec M_0(\vec r)$, the result is 
\be G\approx F[\vec
M_0]+\frac12\log\det\left.\left(\frac{\delta^2F}{\delta\vec M\delta
      \vec M}\right)\right|_{\vec M_0},\label{sp}
\ee 
where we have omitted an additive constant. The first term of the right-hand side is the mean-field contribution $G_{\mbox{\scriptsize mf}}$ and the second term, $G_{\mbox{\scriptsize fluct}}$, is the correction from Gaussian fluctuations. If corrections from fluctuations become of order 1 in the Ginzburg regime close to $T_c$, the saddle point approximation breaks down and Eq.~(\ref{sp}) is no longer valid.

By choosing appropriate units, we can eliminate two of the four parameters in Eq.~(\ref{F}).  For this, we re-scale all lengths as $\tilde{\vec r}=Q \vec r$, magnetizations as $\tilde{\vec M}=[U/(JQ^2)]^{1/2}\vec M$, and fields as $\tilde B=[U/(JQ^2)^3]^{1/2}\vec B$  and obtain
\be F=\gamma\int\!d^3\tilde r\,\left[ (1+t)\tilde{\vec M}^2+(\tilde\nabla\tilde{\vec
    M})^2+2\,\tilde{\vec M}\cdot(\tilde\nabla\times\tilde{\vec
    M})+\tilde{\vec M}^4 -\tilde\vec B\cdot \tilde{\vec M}\right], \label{tF}
\ee 
where $t=r_0/(JQ^2)-1\propto T-T_c^{MF}$ measures the distance to the mean-field critical temperature $T_c^{MF}$ (i.e. within the saddle-point approximation, the system is spiral spin-ordered for $t<0$ and paramagnetic for $t>0$) and $\gamma=J^2Q/U$ provides a relative weight between the mean-field and fluctuation
contributions to Eq.~(\ref{sp}). Therefore, the physics of the model (\ref{F}) depends on three parameters: $\gamma$, $t$ and the magnetic field. From now on, we will omit the tildes in most formulas to simplify notation, but keep in mind that we have chosen particular units.

\subsection{Intuitive arguments for the skyrmion lattice solution}
\label{intuition}

It is possible to give simple geometric arguments, which lead to the specific spin crystal advertised in this paper. We may single out the ferromagnetic component of the magnetic structure $\vec M_f=\int \vec{M}(\vec{r}) d^3 \vec{r}/V$, where $V$ is the volume. The ferromagnetic component $\vec M_f||\vec B$ is induced by the external field. The true order parameter is therefore $\vec \Phi=\vec M-\vec M_f$. The quadratic part is minimized by a helix with wave length $2\pi/Q$ described by 
\be \label{helix} \vec{M}^h_{\vec{Q}}(\vec{r})=A
(\vec{n}_{1} \cos[\vec{Q} \vec{r}]+\vec{n}_{2} \sin[\vec{Q} \vec{r})], 
\ee
where the wave vector $\vec{Q}$ and two unit vectors, $\vec{n}_{1},\vec{n}_{2}$ are orthogonal to each other. The sign of $D$  in Eq.~\ref{F} determines the chirality with $\vec{Q}\cdot (\vec{n}_{1} \times \vec{n}_{2})>0$ for $D>0$.  To quadratic order, also arbitrary linear combinations of such helices minimize the free energy and only the interaction  $\vec{M}^4$ selects the magnetic structure. Expanding this term, we obtain 
\be \vec{M}^4=\vec
M_f^4+4 \vec{M}_f^2 \, \vec{\Phi}\cdot \vec M_f +2\vec M_f^2\vec{\Phi}^2+4(\vec
M_f\cdot\vec{\Phi})^2+4\vec{\Phi}^2\,\vec{\Phi}\cdot \vec M_f+\vec{\Phi}^4 
\ee 
The crystaline state can gain energy from the second last term $\propto \vec{\Phi}^2\,\vec{\Phi}\cdot \vec M_f$, which is cubic in the order parameter.  Fourier transformation leads to
\begin{equation}
\int \vec M_f\cdot\vec{\Phi}\, \vec{\Phi}^2\, d^3\vec{r}=\sum_{\vec{q}_1,\vec{q}_2,\vec{q}_3\neq 0}
(\vec{M}_f\cdot \vec{m}_{\vec{q}_1}) (\vec{m}_{\vec{q}_2}\cdot \vec{m}_{\vec{q}_3})
\delta(\vec{q}_1+\vec{q}_2+\vec{q}_3),\label{cubic}
\end{equation}
where $\vec{m}_{\vec{q}}$ is the Fourier transform of $\vec{M}(\vec r)$.  Therefore, the cubic term vanishes unless the magnetic structure contains Fourier modes of at least three wavevectors $\vec Q_j$ with $\vec Q_1+\vec Q_2+\vec Q_3=0$. If in addition, $|\vec Q_j|=Q$, we immediately arrive at the coplanar arrangement of
Fig.\,3A of the main text. Thus we obtain six Bragg spots with 60 degrees angles between them. Direct calculation shows that in this situation, the vector $ \int \vec{\Phi}^2\,\vec{\Phi}$ is orthogonal to the plane containing $\vec Q_1,\vec Q_2,\vec Q_3$ with a prefactor that depends on the relative phases of the three
helices. Hence, the cubic term is minimized if $\vec Q_1,\vec Q_2,\vec Q_3$ are orthogonal to $\vec B$ (the relative phases then ensure that $ \int \vec{\Phi}^2\,\vec{\Phi}$ becomes anti-parallel to $\vec M_f$, which minimizes the energy gain by the cubic term). Hence, minimization of the cubic term already explains the experimentally observed Bragg spots with hexagonal symmetry in the plane orthogonal to the field.

We have also checked quantitatively that the cubic term plays an essential role in the energetics of the crystal phase. Namely, if the cubic term is artificially put to zero in the evaluation of the free energy,  the crystal state no longer comes energetically close to the conical phase at intermediate fields.

\subsection{Calculation of the mean-field free energy in an external field}
\label{mean-field}

The mean-field ground state (i.e. the global minimum of the functional $F[\vec{M}]$) can be found rigorously. For $B<B_{c2}=(-2t)^{1/2}$, it is the well-known \cite{izyu84} conical phase $\vec{M}(\vec{r})=m \hat{z}+a (\hat{x} \cos(z)+\hat{y} \sin(z))$ with $m=B/2$, $a=\frac{1}{2}\sqrt{-2 t-{B}^2}$ and with the free energy 
\be
\frac1V G_{\mbox{\scriptsize mf}}^{\mbox{\scriptsize con}}=-\frac\gamma4(t^2+\vec B^2), 
\ee 
where $V$ is the system volume.  The conical phase evolves continuously into the helical state at zero field. For  larger fields, $B>B_{c2}$, the ferromagnetic state  becomes the ground state.  The proof is provided most easily by writing $F[\vec{M}]$ in the following way: 
\be \frac1{\gamma }F=-V \frac{t^2+\vec B^2}{4}+V \sum_{\vec q\neq\vec 0}\vec
m_{-\vec q}^\alpha\left[r^{\alpha\beta}(\vec q)-t\delta^{\alpha\beta}\right]\vec m_{\vec q}^\beta+\int
\left(\vec M^2+\frac t2\right)^2 d^3 {r}+V \left({M}_f-\frac{\vec B}2\right)^2,\label{proof} \ee
where $r^{ab}(\vec{k})=(1+t+k^2)\delta^{ab}-2i\epsilon^{abc}k^c$. The conical phase manages to minimize each term of the right-hand side individually.

However, as argued above, there is a competing state at intermediate fields, which we call the A-crystal. This state can be easily addressed in the limit of small $t<0$, where it is obtained from a superposition of the uniform component $\vec M_f$ with three left-handed helical spin spirals. More precisely, 
\be \vec M(\vec
r)=\vec M_f+\sum_{j=1}^3 \left(\vec m_{\vec Q_j}e^{i \vec Q_j\cdot\vec
    r}+\mbox{c. c.}\right), 
\ee 
where $\vec Q_1, \vec Q_2, \vec Q_3$ are wavevectors orthogonal to the field, with $|\vec Q_j|=Q$ and mutual angles of $120^\circ$ between them as shown in Fig.\,3A of the main text, "c.~c." denotes the complex conjugate and 
\be \vec m_{\vec
  Q_j}=\frac12\psi_j(\uv\epsilon'_{\vec Q_j}+i \uv\epsilon''_{\vec
  Q_j}),
\ee 
where $\uv\epsilon'_{\vec Q_j}$, $ \uv\epsilon''_{\vec Q_j}$ are orthogonal unit vectors with $\uv\epsilon'_{\vec Q_j}\times \uv\epsilon''_{\vec Q_j}=\uv Q_j$ and $\psi_j$ is a complex number which encodes the amplitude and phase of the helical spin-density wave $j$. Thus, 
\be \vec M(\vec r)=\vec M_f+\sum_{j=1}^3|\psi_j|(\cos( \vec Q_j\cdot\vec r+\phi_j)\uv\epsilon'_{\vec Q_j}-\sin( \vec
Q_j\cdot\vec r+\phi_j)\uv\epsilon''_{\vec Q_j}), \label{crystal} 
\ee 
where $\psi_j=|\psi_j|\exp{(i \phi_j)}$. Higher order Fourier modes vanish in the limit of small $-t$ (but are included in our numerical calculations for finite $t$). The minimization process leads to $|\psi_1|=|\psi_2|=|\psi_3|$, i.e. the three helices have equal weight. In addition, minimization also fixes the relative phases of the three helices in such a way that at one point (say $\vec r=\vec 0$), the magnetization of each helix points opposite to the field direction. This point is the center of the anti-skyrmions, shown in Fig.\,3C of the main text. The values of the relative phases have
a strong influence on the resulting magnetic structure and  determine whether one obtains a lattice of skyrmions as described in the main text.

\begin{figure}
\begin{center}
\includegraphics[width=.65\textwidth,clip=]{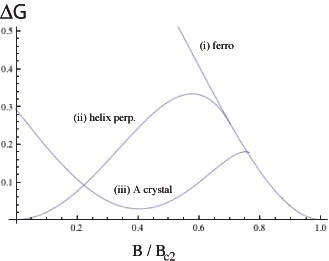}
\end{center}
\caption{Difference of mean-field free energy between various magnetic states and the conical state as a function of magnetic field. The energy difference is plotted in units of $\gamma t^2/4$ in the limit of small $t<0$.} \label{figure-S2}
\end{figure}

In Fig.~\ref{figure-S2}, we plot 
\be \frac 4{\gamma\, t^2V}\Delta G_{\mbox{\scriptsize mf}}=\frac 4{\gamma\,
t^2V}\left(G_{\mbox{\scriptsize mf}}-G_{\mbox{\scriptsize mf}}^{\mbox{\scriptsize con}}\right) 
\ee 
as a function of field in the limit of small $t<0$. The energy difference to the conical phase is therefore plotted in units of $\gamma\, t^2/4$ which corresponds to the energy difference between ferromagnetic and conical states at zero field. For comparison, we have also included (i) the energy of a single helix oriented orthogonal to the field, which corresponds to the traditional interpretation of the  A-phase, and (ii) the energy of the ferromagnetic state. The free energy difference between the A-crystal and the conical state obtains its minimal value $(4/\gamma\,t^2V)\Delta G=0.02838 $ at an external field of $B=0.40357 B_{c2}$. At this point, the A-crystal
obeys $M_f=0.7773|\psi_j|=B/2$.

\begin{figure}
\begin{center}
\includegraphics[width=.65\textwidth]{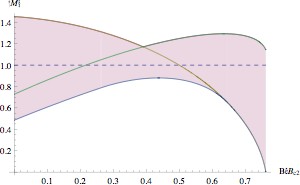}
\end{center}
\caption{Variation of magnetization amplitudes as a function of field, in the limit of small $t<0$. The gray region shows the range of $|\vec M(\vec r)|$ inside the unit cell of the A-crystal, plotted in units of $(-t/2)^{1/2}$.   The two crossing lines correspond to the amplitudes at two specific points in the crystal. For the decreasing line, this is the center of an anti-skyrmion (where $\vec M$ points in the opposite field direction) and for the increasing line the mid-point between three anti-skyrmions (where $\vec M$ points along the field). The dashed line shows the amplitude of the moment in the conical phase, which is a constant for all fields. Note that the A-crystal is most favorable, when the variation of amplitudes is \textit{smallest} rather than largest. } \label{figure-S3}
\end{figure}

As can be seen in Fig.~\ref{figure-S3},  the field where the energy of the A-crystal and the conical phase are closest, is characterized by small spatial variations of the magnetization amplitude $|\vec M|$. This observation reflects the fact that it is energetically unfavourable to have too strong modulations of the magnetization amplitude. In particular, it would cost a large energy to suppress $\vec M$ locally to zero, which makes it meaningful to classify magnetic configurations into topological sectors. For  example, we find that the A-crystal and the conical phase belong to different topological sectors, since both configurations cannot be continuously transformed into each other without suppressing  $\vec M$ locally to zero. In contrast, other non-trivial spin textures which have been proposed earlier in the context of noncentrosymmetric magnets \cite{binz06,roes06,fisc08}, all have points or lines with vanishing magnetization, making any notion of topological stability ill-defined.

Equation \eref{crystal} gives the exact form of the A-crystal in the limit $t\to 0$, very close to the critical temperature. But for $t<0$, the crystal obtains additional corrections, which further suppress amplitude modulations. Namely, the spin structure obtains Fourier weight from all wavevectors of the reciprocal lattice $n\vec Q_1+m\vec Q_2$ and each wavevector is composed of two transversal modes (left- and right-handed helices) and a longitudinal one (where $\vec m_{\vec Q}\parallel\vec Q$). To calculate the mean-field energy, we include wavevectors up to a short-distance cutoff, $|\vec Q|\leq\Lambda$, and minimize numerically with respect to amplitudes and phases of all modes. The structure of the A-crystal gets therefore distorted from the simple form of Eq. \eref{crystal}, but without changing its symmetry or topology.

In our mean-field Ginzburg-Landau theory, the A-crystal thus appears as a metastable phase, which for intermediate fields becomes remarkably close in energy to the conical phase. Experiment suggests that close to the critical temperature, the crystal state becomes {\it stable} for these fields. In the following Section, we show that fluctuation corrections to the mean-field theory readily explain the stability of the A-crystal within the model (\ref{F}).

\subsection{Fluctuation contribution to the free energy}
\label{fluctuations}

We now consider the effect of fluctuation contributions to Eq.~(\ref{sp}): \be G_{\mbox{\scriptsize fluct}}=\frac12\log\det \left(g^{ab}_{\vec{k}\vec{k'}}\right), \ee where $g$ is the matrix of second derivatives,
\begin{eqnarray}
g^{ab}_{\vec{k}\vec{k'}}&=&\frac12\frac{\partial^2F}{\partial
  m^a_{-\vec{k}}\partial m^b_{\vec{k'}}}\\
&=&\gamma\left(\delta_{\vec{k},\vec{k'}}r^{ab}(\vec{k})+2\delta^{ab}\sum_{\vec{k''}}\vec{m_{-\vec{k''}}}\cdot\vec{m_{\vec{k-k'+k''}}}+4\sum_{\vec{k''}}m^a_{-\vec{k''}}m^b_{\vec{k-k'+k''}}\right).\label{g}
\end{eqnarray}
We will be interested in the energy difference between the conical phase and the A-crystal 
\be 
\Delta G_{\mbox{\scriptsize fluct}}=G^{\mbox{\scriptsize A}}_{\mbox{\scriptsize
    fluct}}-G^{\mbox{\scriptsize con}}_{\mbox{\scriptsize fluct}}.
\ee

Surprisingly, the energy difference between conical phase and A-crystal obtains a contribution even from short length-scales as we show in the following. First, we write $g^{ab}_{\vec{k}\vec{k'}}= g_0+g_1$, where $g_0$ is the first term on the right-hand side of Eq.~(\ref{g}), which is independent of the magnetic state, and $g_1$ are the two remaining terms, which are quadratic in $\vec{m}_{\vec{k}}$. For large values of $\vec{k}$ or  $\vec{k'}$, the term  $g_0$ behaves asymptotically like $g_0\approx\delta^{\alpha\beta}\delta_{\vec{k k'}}\, k^2$ and dominates over $g_1$. The short-wavelength contribution can therefore be expanded as 
\be
 G_{\mbox{\scriptsize fluct}}=\frac1{2}\left(\tr\log
  g_0+\tr(g_0^{-1}g_1)+\tr(g_0^{-1}g_1g_0^{-1}g_1)+\ldots\right).
\ee We finally obtain the short-wavelength contribution to the energy difference as \be \frac1V\Delta
G_{\mbox{\scriptsize fluct}}^{\mbox{\scriptsize
    short}}=\frac{5\Lambda}{2\pi^2}\left[
\frac t2+ \frac{1}{V} \int \vec{M}^2_A \right], 
\ee where $\Lambda$ is an ultraviolet wavevector cutoff, $\vec{M}_A(\vec{r})$ is the magnetization  in the A-crystal and we have used the fact that $\vec{M}^2=-t/2$ in
the conical phase. In the presence of higher-order gradient terms in $g_0$ one can use the equation $\int g_0 d^3k/(2 \pi)^3=\Lambda/(2 \pi^2)$ as a definition of $\Lambda$.

The remaining part 
\be \Delta G_{\mbox{\scriptsize fluct}}^{\mbox{\scriptsize long}}=\Delta G_{\mbox{\scriptsize fluct}}-\Delta G_{\mbox{\scriptsize fluct}}^{\mbox{\scriptsize short}} 
\ee 
has a finite limit for $\Lambda\to\infty$ and needs to be calculated numerically. For periodic structures such as the conical phase or the A-crystal, one may decompose $\vec{k}=\vec{Q}+\vec{q}$, where $\vec{Q}$ is a reciprocal lattice vector and
$\vec{q}$ is in the first Brillouin zone. The fluctuation matrix is then of the form
$g^{ab}_{\vec{k}\vec{k'}}=\delta_{\vec{q,q'}}\,g^{ab}_{\vec{Q},\vec{Q'}}(\vec{q})$, such that \be \frac1V \Delta
G_{\mbox{\scriptsize
    fluct}}=\frac1{2}\int_{\mbox{\scriptsize
    B.Z.}}\!\frac{d^3q}{(2\pi)^3}\,\log\frac{\det
  \left[\left.g^{ab}_{\vec{Q},\vec{Q'}}(\vec{q})\right|_{\mbox{\scriptsize
     A}}\right]}{\det
  \left[\left.g^{ab}_{\vec{Q},\vec{Q'}}(\vec{q})\right|_{\mbox{\scriptsize
      con}}\right]},
\ee where the integral is over the Brillouin zone. In principle, we could chose a one-dimensional reciprocal lattice for the conical phase and a two-dimensional one for the A-crystal. However, in order to treat the two phases on equal footings, we chose a  three-dimensional reciprocal lattice spanned by $\vec{Q}_1,\vec{Q}_2$ from Fig.~3A  in the main text perpendicular to the field and a third vector of the  same length along the field. Thus, the Brillouin zone obtains the shape of a hexagonal  prism. We introduce the ultraviolet cutoff by truncating the matrix at $|\vec{Q}|,|\vec{Q'}|\leq \Lambda$.

To obtain the correct  Gaussian fluctuations it is important to use the numerically exact saddle-point within the chosen cutoff. Therefore, we can not restrict it to the simple form of Eq.~(\ref{crystal}) for the A-crystal but perform a numerical minimization with respect to all modes $\vec{m}_{\vec{Q}}$ for $|\vec{Q}|\leq\Lambda$ to determine the saddle point. We also minimize with respect to the fundamental helix wavevector length $Q$, which turns out to be slightly reduced from unity.

\begin{figure}
\begin{center}
\includegraphics[width=.85\textwidth]{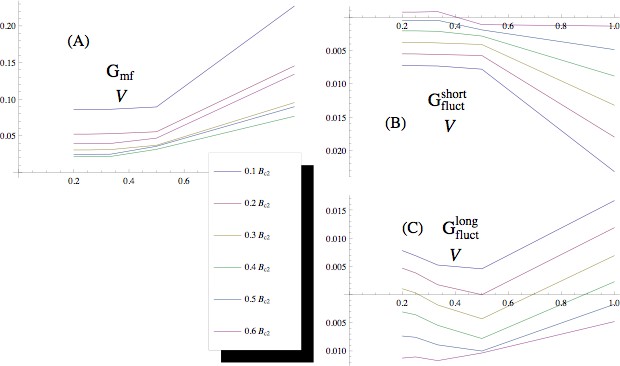}
\end{center}
\caption{The three contributions to $\Delta G=\Delta G_{\mbox{\scriptsize mf}}+\Delta G_{\mbox{\scriptsize
fluct}}^{\mbox{\scriptsize
    short}}+\Delta G_{\mbox{\scriptsize fluct}}^{\mbox{\scriptsize
    long}}$, plotted as a function of $1/\Lambda$ for $t=-4$ and various fields. The energies per unit volume have been rescaled by $\gamma$ or $\Lambda$ as indicated in the figures. Numerical calculations have been performed up to $\Lambda=5$. The data can be readily extrapolated to the physical cutoff $\Lambda\approx 40$ (in units of $Q$). Similar results are obtained for smaller values of $|t|$.
For much larger $|t|$, one would have to increase the cutoff to obtain convergence, which makes the numerical computations more demanding.} \label{figure-4S}
\end{figure}

After the minimization, which determines $\Delta G_{\mbox{\scriptsize mf}}$, we calculate $\int \vec{M}^2$ and obtain $\Delta G_{\mbox{\scriptsize fluct}}^{\mbox{\scriptsize short}}$. Then, we calculate the fluctuation matrices and evaluate the Brillouin zone integral numerically, to determine  $\Delta G_{\mbox{\scriptsize fluct}}^{\mbox{\scriptsize long}}$. In Fig.~\ref{figure-4S}, we show, how the different contributions to $\Delta G=\Delta G_{\mbox{\scriptsize
mf}}+\Delta G_{\mbox{\scriptsize fluct}}^{\mbox{\scriptsize
    short}}+\Delta G_{\mbox{\scriptsize fluct}}^{\mbox{\scriptsize
    long}}$ depend on the cutoff. These results confirm that   $\Delta
G_{\mbox{\scriptsize fluct}}^{\mbox{\scriptsize long}}$ has indeed a finite limit for $\Lambda\to\infty$. These figures also show that fluctuation contributions at both short and long length-scales favor the A-crystal over the conical phase. Due to the thermal fluctuations the A-crystal obtains energetic stability as shown in the inset of Fig.~3B in the main text, where $\Delta G_{\mbox{\scriptsize mf}}$ and $\Delta G$ are shown in units of $\gamma t^2/4$.

In real systems, the cutoff is of the order of $2\pi/a$, where $a$ is the lattice constant. For the case of MnSi, we have $\Lambda\approx 40 Q$ (where $Q=1$ in our units). For such large cutoff values, the short-distance contributions dominate over the long-distance ones, such that \be \Delta G\approx\Delta G_{\mbox{\scriptsize
    mf}}+\frac{5\Lambda}{2\pi^2}\left[\frac
  t2+\frac{1}{V} \int \vec{M}^2_A \right]\label{approx}
\ee is a very good approximation except for extremely small values of $|t|$ (the asymptotic small-$t$ behavior is $\Delta G_{\mbox{\scriptsize mf}}\propto t^2$, $\Delta G_{\mbox{\scriptsize
    fluct}}^{\mbox{\scriptsize short}}\propto t$ and  $\Delta G_{\mbox{\scriptsize
    fluct}}^{\mbox{\scriptsize long}}\propto \sqrt{-t}$). This approximation was used to establish the phase diagram shown in Fig.\,3B in the main text.

Interestingly, the A-crystal is even no longer a local minimum of $F$ for too small values of $B$, i.e., below the red dotted line in Fig.~3B in the main text. There, it becomes locally unstable (i.e. unstable with respect to small variations) since the fluctuation matrix $\delta^2F/(\delta\vec M)^2$ obtains a negative eigenvalue. However, we find that the A-crystal at intermediate fields is both locally stable and energetically favorable to the conical phase due to fluctuation corrections to the free energy.

\subsection{Testing the validity of the saddle-point approximation}
\label{testing}

The saddle-point approximation, as expressed in Eq.~\eref{sp}, is obtained by an expansion around a local minimum of $F$. It is thus only valid, if fluctuations around the minimum are small or, technically, if fluctuation corrections to the size of the order parameter are small compared to its mean-field value. To test this hypothesis, we calculate the response with respect to a fictitious space-dependent field $\vec B(\vec r)$ and obtain the local magnetization as $\vec M(\vec r)=-{\delta G}/{\delta \vec B(\vec r)}$. Using Eq.~\eref{sp}, we obtain 
\be M^j=M^i_0-\left.\frac18\chi^{i_1i_2}\frac{\delta^3F}{\delta
    M^{i_1}\delta M^{i_2}\delta M^{i_3}}\chi^{i_3j}\right|_{\vec M_0},\label{opcorr1}
\ee 
where $M^i=M^\alpha(\vec r)$ (i.e. $i,j$ indices stand for both a space coordinate as well as a directional index) and $\chi^{-1}=1/2\delta^2F/(\delta M)^2$. The second term on the right-hand side is the fluctuation-correction to the order parameter. If the saddle-point approximation is valid, we expect this correction to be small. As a test for the validity of our approximation, we therefore demand that the second term in Eq.~\eref{opcorr1} is small compared to the first one.

More precisely, we take the Fourier transform of Eq.~\eref{opcorr1} and obtain the fluctuation-correction to the order parameter as 
\be \delta m_{\vec Q}^a=-\!\!\!\sum_{\vec Q_1,\vec Q_2,\vec Q_3}\int_{\mbox{\scriptsize
    B.Z.}}\!\frac{d^3q}{(2\pi)^3}\,\chi_{\vec Q_1\vec
  Q_2}^{a_1a_2}(\vec q)\left(\delta_{a_1a_2}m^{a_3}_{\vec Q_1-\vec Q_2+\vec Q_3}+2\delta_{a_2a_3}m^{a_1}_{\vec Q_1-\vec Q_2+\vec Q_3}\right)\chi_{\vec Q_3\vec Q}^{a_3a}(\vec 0),\label{opcorr2}
\ee 
where $\chi_{\vec Q\vec Q'}^{ab}(\vec q)$ is the inverse matrix of $g_{\vec Q\vec Q'}^{ab}(\vec q)$. For the saddle-point approach to be valid, we demand that \be |\delta \vec m_{\vec Q}|\ll  |\vec m_{\vec Q}| \ee for the dominant modes with $|\vec Q|=1$.

In the same way as above for the fluctuation-correction to the free energy, it turns out that Eq.~\ref{opcorr2} obtains a cutoff-dependent contribution from short length-scales as follows: 
\be \delta m_{\vec Q}^a=-\frac{3\Lambda}{2\pi^2\gamma}\sum_{\vec   Q',b}m_{-\vec Q'}^b\chi^{ba}_{\vec Q'\vec Q}(\vec 0)+O(\Lambda^0).\label{opcorrapprox} 
\ee 
For large values of the cutoff, such as $\Lambda\approx 40$, the short-distance part ($\propto \Lambda$) dominates over the long-distance part [$O(\Lambda^0)$]. We therefore use Eq.~\eref{opcorrapprox} to map out the regime of validity, which we define as the region, where $|\delta \vec m_{\vec Q}|<0.2 |\vec m_{\vec Q}|$, in the phase diagram [red dashed line in Fig.~3B in the main text].

\subsection{Pinning of spin crystal to the atomic lattice}
\label{pinning}

In our experiment we observe that the six-fold diffraction pattern of the A-phase gets weakly oriented with respect to the atomic crystal within the plane perpendicular to the magnetic field. The pinning to the crystal is such that  two of the six peaks point in a $\langle 110\rangle$ crystal direction, as long as this is compatible with the field orientation. Here we show that this behavior is fully compatible with our phenomenological theory of the A-crystal.

The model of Eq. \eref{F} is fully rotation symmetric. A rotation of the magnetic structure is performed by $\vec M(\vec r)\to R \vec M(R^{-1}\vec r)$, where $R$ is an $SO(3)$ operator. The coupling to an external field breaks the rotation symmetry to $SO(2)$, the group of rotations around the field-direction. Terms which are higher-order in spin-orbit coupling (neglected in Eq. \eref{F}) will break this symmetry and couple the magnetic structure to the underlying atomic lattice.

To investigate pinning systematically, we organize these terms in powers of spin-orbit coupling \cite{foot1}. In our mean-field Ginzburg-Landau approach, this means $t\to0$ and $M\sim B^*\sim (-t)^{1/2}$. The energy scale of helix formation is of second order in spin-orbit coupling\cite{naka80,bak80}. The leading anisotropy terms are of fourth order  and given by \cite{foot2} 
\be \Delta F^{(4)}=\lambda_1\sum_{q}(\hat q_x^4+\hat q_y^4+\hat q_z^4)|\vec m_{\vec q}|^2+\lambda_2 \int M_x^4+M_y^4+M_z^4.\label{Fa} 
\ee 
We treat these anisotropy terms perturbatively, i.e., we first determine the magnetic structure using Eq. \eref{F} and then calculate how $\Delta F^{(4)}$ changes if the structure is rotated around the field direction. This gives a pinning function $F(\varphi)$, where $\varphi$ is the rotation angle relative to some standard orientation.

The leading anisotropy term  of  Eq. \eref{Fa} provides the pinning of the single helix state in MnSi and related helimagnets. In MnSi, the parameters $\lambda_1,\lambda_2$ are such that $\hat q_x^4+\hat q_y^4+\hat q_z^4$ is minimized, i.e., the preferred spiral direction at zero field is $\langle 111\rangle$. If we assume for a moment, that the A-phase consists of a single spiral which is oriented orthogonal to the field, this spiral would lock into a direction $\uv q$ which minimizes $\hat q_x^4+\hat q_y^4+\hat q_z^4$ under the constraint that $\uv q\cdot \uv B=0$. This scenario is not compatible with the observations from neutron scattering (Fig.\,2 in the main text).

Interestingly, the anisotropy term in Eq. \eref{Fa} does {\it not} lead to a pinning of the hexagonal spin crystal to linear order in $\lambda_1$ and $\lambda_2$ (i.e. to fourth order in spin-orbit coupling). The reason is the six-fold rotation axis of this crystal. As a consequence, the pinning function is of the form  $F(\varphi)=\sum_n(a_n \cos 6n\varphi+b_n \sin 6n\varphi$), but the terms of Eq. \eref{Fa} are not capable of producing so rapidly oscillating functions to linear order. The pinning potential of the spin crystal state is therefore produced by  next order terms (sixth order in spin-orbit coupling), for example by 
\be 
\Delta F^{(6)}=\lambda_3\sum_{q}\hat q_x^2\hat q_y^2\hat q_z^2\,|\vec m_{\vec q}|^2. 
\ee 
We have analyzed the pinning of the spin crystal in the presence of $\Delta F^{(6)}$. We obtain for $\lambda_3>0$ that the A-crystal gets oriented in such a way that at least two of the six Bragg spots point along a $\langle 110\rangle$ direction if the field direction allows for it (i.e., if the field is orthogonal to one of the   $\langle 110\rangle$
directions). This is compatible with the observations shown in Fig.\,2 of the main text.

\newpage



\newpage

\vspace{5mm}
\newpage

\vspace{5mm}

\newpage

\newpage

\vspace{5mm}

\newpage

\newpage

